\documentclass[%
reprint, twocolumn,
superscriptaddress,
nofootinbib,
amsmath,amssymb, aps, pra,
]{revtex4}

\usepackage{amsmath}
\usepackage{amssymb}
\usepackage{graphicx}
\usepackage{dcolumn}
\usepackage{bm}
\usepackage{color} 
\usepackage{CJK}
\usepackage{graphicx}
\usepackage{subfigure}
 \usepackage{mathrsfs}
 \usepackage{dsfont}
  \usepackage{nicefrac}
\usepackage[extension=xxx]{hyperref}

\usepackage{lipsum}

\newcommand {\fabs}[1] {\left| #1 \right|}

\newcommand {\fabsq}[1] {\left| #1 \right|^2}
\newcommand {\la}{\langle}
\newcommand {\ra}{\rangle}

\newcommand{\ket}[1]{\ensuremath{|#1\rangle}}
\newcommand{\bra}[1]{\langle#1|}
\newcommand{\braket}[2]{\langle#1|#2\rangle}
\newcommand{\ketbra}[2]{|#1\rangle\langle#2|}

\def\la{{\langle}}
\def\ra{{\rangle}}

\def\wh{\widehat}
\def\nn{\nonumber}
\def\da{^\dagger}

\newcommand{\beq}{\begin{equation}}
\newcommand{\eeq}{\end{equation}}
\newcommand{\beqa}{\begin{eqnarray}}
\newcommand{\eeqa}{\end{eqnarray}}

\begin{document}
\title{$S$-matrix pole symmetries for non-Hermitian scattering Hamiltonians}
\author{M. A. Sim\'on}
\affiliation{Department of Physical Chemistry, Universidad del Pa\'{\i}s Vasco - Euskal Herriko Unibertsitatea, Apdo. 644, Bilbao, Spain}
\author{A. Buend\'ia}
\affiliation{Department of Physical Chemistry, Universidad del Pa\'{\i}s Vasco - Euskal Herriko Unibertsitatea, Apdo. 644, Bilbao, Spain}
\author{A. Kiely}
\affiliation{Department of Physical Chemistry, Universidad del Pa\'{\i}s Vasco - Euskal Herriko Unibertsitatea, Apdo. 644, Bilbao, Spain}
\affiliation{Department of Physics -University College Cork, Ireland}
\author{Ali Mostafazadeh}
\affiliation{Departments of Mathematics and Physics, Ko\c{c} University,
34450 Sar\i yer, Istanbul, Turkey}
\author{J. G. Muga}
\affiliation{Department of Physical Chemistry, Universidad del Pa\'{\i}s Vasco - Euskal Herriko Unibertsitatea, Apdo. 644, Bilbao, Spain}
\begin{abstract}
%

The complex eigenvalues of some non-Hermitian Hamiltonians,
e.g. parity-time symmetric Hamiltonians, come in complex-conjugate pairs. We show that for non-Hermitian scattering Hamiltonians (of a structureless particle in one dimension) possesing one of four certain symmetries, the poles of the $S$-matrix eigenvalues in the complex  momentum plane are symmetric about the imaginary axis, i.e. they  are complex-conjugate pairs in complex-energy plane. This applies even to states which are not bounded eigenstates of the system, i.e. antibound or virtual states, resonances, and antiresonances. The four Hamiltonian
symmetries are formulated as the commutation of the Hamiltonian with specific antilinear operators. Example potentials with such symmetries are constructed and their pole structures and scattering properties are calculated.
\end{abstract}
\maketitle
\section{Introduction}
\label{sec:intro}
Non-Hermitian (NH) Hamiltonians may represent effective interactions for components of a system. Feshbach's partitioning technique \cite{Feshbach1958,Feshbach1962} provides the formal framework to find NH-Hamiltonians for a subspace, from the Hermitian Hamiltonian for the total system. NH-Hamiltonians are also set phenomenologically to mimic some observed or desired behaviour, such as gain, decay or absorption in nuclear or atomic, molecular, and optical physics \cite{Muga2004,Ge2017,ChristodoulidesNat2018}. They arise as well as auxiliary tools to facilitate calculations of cross sections  or resonances, e.g. by complex scaling of the coordinates \cite{ComplexScaling1,ComplexScaling2}, and also to model some open systems \cite{rotter2009non} and lattices \cite{alvarez2018topological}.

Much work on  NH-physics  has focused on PT-symmetric Hamiltonians,
as they may have a purely real spectrum \cite{BenderBoetcher1998}. More recently, other NH and non-PT Hamiltonians, have been shown to hold real eigenvalues \cite{Nixon2016,Chen2017,Yang2017}. Work on scattering by PT-symmetric potentials was at first rather scarce
 \cite{Muga2004,Ruschhaupt2005,Cannata2007,Znojil2015}. However, scattering has been later investigated intensely
in connection with spectral singularities and reflection asymmetries for left or right incidence (i.e. unidirectional invisibility)
 \cite{Mosta2009,Longhi2014,Mosta2013}, in most cases restricting the analysis to local potentials.
It has been recently shown  that different devices with asymmetrical scattering responses (i.e., with different transmission and/or reflection for right and left incidence in a 1D setting) are possible if one makes use of non-local potentials \cite{Ruschhaupt2017}.
Ref. \cite{Ruschhaupt2017} provides the selection rules for the transmission and reflection coefficient asymmetries based on eight basic Hamiltonian symmetries. Four of these symmetries are of the standard form,
\beq
AH=AH,
\label{gs1}
\eeq
and the other four are of the $A$-pseudo-hermitian form
\beq
AH=H^\dagger A,
\label{gs2}
\eeq
where $A$ is a unitary or anti-unitary operator in Klein's $4$-group $\mathbf{K}_4 = \left\{1,\Pi,\Theta,\Theta\Pi\right\}$ formed by the identity ($1$), parity ($\Pi$), time-reversal ($\Theta$) and their product ($\Theta\Pi$), also termed $PT$.

\begin{figure}[h]
    \includegraphics[width=0.75\linewidth]{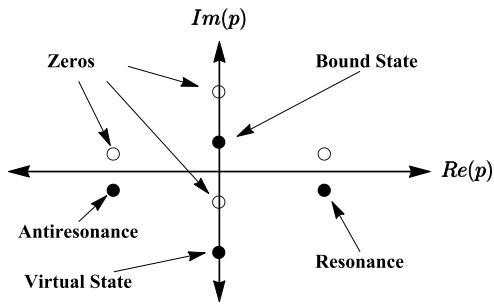}
    \caption{Example of configuration of  poles (filled circles) and zeros (empty circles) of the $S$-matrix eigenvalues in the complex momentum plane for hermitian Hamiltonians.  Poles in the upper half plane ($\operatorname{Im}(p) > 0$) correspond to bound eigenstates of the Hamiltonian, i.e. localized states with negative energy. Poles in the lower half plane correspond to  virtual states ($\operatorname{Re}(p) = 0$), resonances ($\operatorname{Re}(p) > 0$) and antiresonances ($\operatorname{Re}(p)<0$). The singularities with negative imaginary part correspond to states that do not belong to the Hilbert space since they are not normalizable. However, they can produce observable effects in the scattering amplitudes, in particular when they approach the real axis. The pole structure of symmetries IV, V, and VII, see Table I,
is similar, but pole pairs are also possible in the upper half-plane.}
    \label{fig:DiagramPoles}
\end{figure}

Here we aim at extending further our understanding of scattering of a structureless particle by NH-potentials in 1D
by considering general potentials that
are not necessarily diagonal in coordinate representation (i.e., non-local potentials). These
typically arise when applying Feshbach's partitioning technique.
The results of \cite{Ruschhaupt2017} are expanded in several directions:

i) We provide an alternative characterization of the
above-mentioned eight symmetries in terms of the invariance of $H$ with respect to the action of superoperators. We also show that the four symmetries associated with $A$-pseudohermiticity relations (\ref{gs2}) can be formulated as well as the commutativity of $H$ with  certain operators
(linear if $A$ is antilinear, and antilinear if $A$ is linear). This formulation extends earlier results
for Hamiltonians with a discrete spectrum \cite{Mosta2002a,Mosta2002c}.

ii) Moreover,
four of these symmetries imply the same
type of pole structure of $S$-matrix eigenvalues in the complex momentum plane that was found for PT symmetry \cite{Muga2004},
namely, zero-pole correspondence at complex-conjugate points, and poles on the imaginary axis or forming symmetrical pairs with respect to the imaginary axis.
For Hermitian Hamiltonians with scattering (and possibly bound) eigenstates, their $S$-matrix poles are symmetric in the complex momentum plane with respect to the imaginary axis (see Fig. \ref{fig:DiagramPoles}).
In the upper half-plane,  which corresponds to the first energy Riemann sheet, the poles are on the imaginary axis and represent bound states
in the point spectrum of the Hamiltonian. In the lower half-plane they come in symmetrical
resonance and antiresonance pairs, and may also lie on the imaginary axis as ``virtual states''. A further symmetry is the occurrence of a zero
at the complex-conjugate momentum of a given pole. These properties are well known for partial wave scattering by spherical potentials
but also hold for the $S$-matrix eigenvalues in one dimensional scattering \cite{Muga2004}.

For NH-Hamiltonians  the above pole- and pole/zero-symmetries do not hold in general. The point spectrum of $H$ may include, apart from ``ordinary bound states''
on the imaginary axis, also eigenvalues ($S$-matrix poles)  in the first and second quadrant.
However the symmetries of zeros and poles, which are characteristic of Hermitian potentials are partly or even completely recovered in some special cases.
In particular, parity-time (PT) symmetric scattering potentials were shown to
keep the zero-pole symmetry \cite{Muga2004}.
PT-symmetry also implies that the  poles (in the upper or lower half-planes) occur as symmetrical  pairs
with respect to the imaginary axis (corresponding to conjugate complex energies)
or lie on the imaginary axis (with corresponding real energy) \cite{Muga2004}.

The configuration with poles located on the imaginary  axis or as symmetrical pairs has some important consequences. In particular, it provides stability of the real energy eigenvalues with respect to parameter variations of the potential. While a simple pole on the imaginary axis can move along that axis when a parameter is changed, it cannot move off this axis (since this would violate the pole-pair symmetry) or bifurcate. The formation of pole pairs occurs near special  parameter values for which two poles on the imaginary axis collide.

The remainder of the article is organized as follows. In section \ref{sec:SymTheory} we review the scattering properties of eight different Hamiltonian symmetries. These symmetries may be characterised as commutativity or pseudohermiticity with respect to four unitary or antiunitary operators forming a Klein $4$-group, or as invariance with respect to the action of eight linear or antilinear superoperators. In section \ref{sec:SPoles} we discuss the physical consequences of the symmetries in the pole structure of the scattering matrix eigenvalues  and hence in the transmission/reflection amplitudes. Four symmetries are shown to lead to complex poles corresponding to real energies or conjugate (energy) pairs.  In section \ref{sep_pot_sec} we exemplify the above with separable potentials exhibiting parity-pseudohermiticity and time-reversal symmetry. These are the two non-trivial symmetries of the four (in the sense that the other two, hermiticity and PT-symmetry, are already well discussed). In section \ref{sec:Conclusions} we discuss and summarize our results.


\begin{table*}
\centering
\scalebox{1}{
\begin{tabular}{ccccccccccc}
Code & Symmetry&  $\la x|V|y\ra$ & ${\cal{L}}$(coord)& $\la p|V|p'\ra$ & ${\cal L}$(momentum) &$\la p|S|p'\ra$ & $T^l$ & $T^r$ & $R^l$& $R^r$
\\
\hline
I & $1H=H1$ &   $\la x|V|y\ra$ & ${ 1}$ &$\la p|V|p'\ra$ &${ 1}$& $\la p|S|p'\ra$ & $T^l$ & $T^r$ & $R^l$ & $R^r$
\\
II & $1H=H^\dagger 1$ &  $\la y|V|x\ra^*$ & ${\cal TC}$& $\la p'|V|p\ra^*$ &${\cal T'C'}$&$\la p|\widehat{S}|p'\ra$ & $\widehat{T}^l$& $\widehat{T}^r$ & $\widehat{R}^l$ & $\widehat{R}^r$
\\
III & $\Pi H=H\Pi$ &  $\la -x|V|-y\ra$ &${\cal I}$&  $\la -p|V|-p'\ra$ &${\cal I'}$&$\la -p|S|-p'\ra$ & $T^r$ & $T^l$ & $R^r$ & $R^l$
\\
IV & $\Pi H=H^\dagger \Pi$ &  $\la -y|V|-x\ra^*$ &${\cal CTI}$& $\la -p'|V|-p\ra^*$ &${\cal C'T'I'}$& $\la -p|\widehat{S}|-p'\ra$ & $\widehat{T}^r$ & $\widehat{T}^l$ & $\widehat{R}^r$ & $\widehat{R}^l$
\\
V & $\Theta H=H\Theta$ &  $\la x|V|y\ra^*$&${\cal C}$& $\la -p|V|-p'\ra^*$ &${\cal I'C'}$& $\la -p'|\widehat{S}|-p\ra$ & $\widehat{T}^r$ & $\widehat{T}^l$ & $\widehat{R}^l$& $\widehat{R}^r$
\\
VI & $\Theta H=H^\dagger\Theta$ &  $\la y|V|x\ra$&${\cal T}$& $\la -p'|V|-p\ra$ &${\cal I'T'}$& $\la -p'|S|-p\ra$ & $T^r$& $T^l$ & $R^l$& $R^r$
\\
VII & $\Theta\Pi H=H\Theta \Pi$ &  $\la -x|V|-y\ra^*$ &${\cal IC}$& $\la p|V|p'\ra^*$ &${\cal C'}$& $\la p'|\widehat{S}|p\ra$ &$\widehat{T}^l$& $\widehat{T}^r$ & $\widehat{R}^r$& $\widehat{R}^l$
\\
VIII& $\Theta\Pi H=H^\dagger \Theta \Pi$ &  $\la -y|V|-x\ra$ &${\cal IT}$& $\la p'|V|p\ra$ &${\cal T'}$& $\la p'|S|p\ra$ & $T^l$ & $T^r$ & $R^r$ & $R^l$
\end{tabular}
}
\caption{Symmetries of the potential based on the commutativity or pseudo-hermiticity of $H$ with the elements of $\mathbf{K}_4$ (second column). Columns 3, 5, and 7 to 11 are to be read as follows: For each symmetry the object in the column is equal to the one in the top row of the column. The relations among potential matrix elements are given in coordinate and momentum representations in the third and fifth columns. In columns 4 and 6, each symmetry is regarded as the invariance of the potential with respect to the transformations represented by superoperators ${\cal L}$ (see Sec. \ref{super}) in coordinate  or momentum  representation.
The fifth column gives the relations they imply in the matrix elements of $S$ and $\widehat{S}$ matrices. The final four columns set the relations for the scattering amplitudes.
\vspace*{.2cm}
\label{table2}}
\end{table*}

\section{Hamiltonian Symmetries}
\label{sec:SymTheory}
Let us first clarify the terminology. Scattering Hamiltonians are here of the form
$H=H_0+V$, where $H_0$ is the kinetic energy operator and $V$ is the potential, which decays fast enough to zero in coordinate representation
so that the usual operators of scattering theory are well defined and the Hilbert space is (biorthogonally) decomposed into a continuum
part with real eigenvalues and a discrete part. See Appendix \ref{sec:ScattFormalism} for a review of the formalism and notation we use.

$V$ is in general non-local, i.e., it does not have the local form $\la x|V|x'\ra=\delta(x-x')V(x)$.
Apart from their generic appearance in Feschbach's partitioning technique, see e.g. \cite{Ruschhaupt2004},
non-local potentials   are quite common in models that discretize the coordinates at specific sites, as in tight-binding models, although here we only consider continuous-coordinate models.

We will now discuss the eight symmetries identified in \cite{Ruschhaupt2017},
which are associated with the two generalized symmetry relations corresponding to commutation with $A$ and $A$-pseudohermiticity \cite{Mosta2002a}, see Eqs. (\ref{gs1},\ref{gs2}).
%
%
We use Roman numeral to label these symmetries as shown in Table I.
Note that a local potential would automatically fulfill symmetry VI but this symmetry does not necessarily imply locality.
For local potentials four of the eight symmetries coincide with the other four \cite{Ruschhaupt2017}. Here we consider general nonlocal potentials where all the eight symmetries are distinct.

The generalization of the symmetry concept to the pair \eqref{gs1} and \eqref{gs2} is in fact quite natural if we take into account that a NH-$H$
has generically different left and right eigenvectors. Given a right eigenstate $\ket{\psi}$ of $H$ with eigenvalue $E$, Eq. (\ref{gs1}) implies that  $A|\psi\ra$ is also a right eigenvector with eigenvalue $E$ or $E^*$, whereas Eq. (\ref{gs2}) implies that $\la \psi|A$ is a left eigenvector of $H$ with eigenvalues $E^*$ or $E$,  for $A$ unitary or antiunitary respectively.

The symmetries which imply the presence of real or complex-conjugate pairs of energy eigenvalues for bound eigenstates
are II, IV,V and VII.
The emergence of these complex-conjugate pairs has been previously discussed in \cite{Mosta2002a,Bender2010} for a general class of diagonalizable Hamiltonians that posses a discrete spectrum. They can be heuristically understood for the symmetries we consider as follows: Symmetry V implies that the Hamiltonian must be real in coordinate space, which would lead to a real characteristic polynomial with real or complex-conjugate roots. Symmetry VII is PT symmetry which is well discussed in the literature as having real or complex-conjugate pairs of eigenvalues \cite{BenderBoetcher1998}. Note also that the matrix elements of PT-symmetric Hamiltonians are real in the momentum representation. More generally, in \cite{Mosta2002c}, it was shown, for diagonalizable Hamiltonians having a discrete spectrum, that $A$-pseudo-Hermiticity for a Hermitian invertible linear operator $A$ is equivalent to the presence of an ordinary symmetry of the form: $BH = HB$ for some antilinear operator $B$ with $B^2 = 1$. Because $B$ is an antilinear operator, the eigenvectors $\ket{E_n}$  of $H$ satisfy
\begin{eqnarray}
H B \ket{E_n}&=&B H \ket{E_n} \nonumber \\
				   &=&E_{n}^{*} B \ket{E_n}.
\end{eqnarray}
Therefore complex eigenvalues $E_n$ come in complex-conjugate pairs. In particular, when $\ket{E_n}$  is an eigenvector of $B$, i.e., $B\ket{E_n} = e^{ib_n} \ket{E_n}$ for some real number $b_n$, we have $E_n \in \mathbb{R}$.
The proof of the equivalence of $A$-pseudo-Hermiticity for linear $A$ and the presence of ordinary antilinear symmetries given in \cite{Mosta2002c} relies on the observation that every diagonalizable Hamiltonian with a discrete spectrum is $\tau$-pseudo-Hermitian for some invertible Hermitian antilinear operator $\tau$, i.e.,
$\tau H = H^\dagger\tau$. This relation together with Eq. (\ref{gs2}) implies $BH = HB$,
if we set $B = A^{-1}\tau$. The same construction applies for the cases where $A$ is a Hermitian antilinear operator, in which case $B$ is a linear operator. In Appendix C we extend this result to the class of scattering potentials of our interest. This shows that symmetries IV and VII can also be expressed as $BH = HB$. A novel aspect uncovered in this paper is that whenever one of  the above-mentioned four symmetries are present not only the complex eigenvalues representing the bound states come in conjugate-complex pairs, but all the complex poles of the $S$-matrix have this property.

\subsection{Superoperator formalism \label{super}}
The eight symmetries listed in Table I may also be regarded as the invariance
of the Hamiltonian matrix with respect to
transformations represented by unitary or antiunitary superoperators ${\cal L}$ \cite{Simon2018} defined by
\begin{eqnarray}
\mathcal{L}(H)=
	\begin{cases}
      A^\dagger H A &  \text{I, III,V,VII} \\
      A^\dagger H^\dagger A &\text{II, IV, VI, VIII}
   \end{cases}.
\end{eqnarray}
%
%
%
This definition of the superoperator action is independent of the representation we use, but its realization
in coordinates or momenta in terms of the operations of complex conjugation, transposition, and inversion is different.
For example, in coordinate representation, these superoperators take the following forms (see the third column in Table I),
\beqa
1 H&=&\int\!\!\int |x\ra \la x|H|y\ra\la y| dx dy,
\nonumber\\
{\cal T} (H) &=&\int\!\!\int |x\ra \la y|H|x\ra\la y| dx dy,
\nonumber\\
{\cal C} (H)&=&\int\!\!\int |x\ra \la x|H|y\ra^*\la y| dx dy,
\nonumber\\
{\cal I} (H)&=&\int\!\!\int |x\ra \la -x|H|-y\ra\la y| dx dy.
\label{defs}
\eeqa
%
Adopting the following inner product for linear operators $F$ and $G$, $\langle\langle F|G\rangle\rangle={\rm{tr}} F^\dagger G$,
we can show that all superoperators ${\cal L}$ are either unitary (for ${\cal L}=1,{\cal T},{\cal I},{\cal TI}$), or antiunitary (for ${\cal L}={\cal C}, {\cal CT},{\cal CI},{\cal CTI}$), i.e.,
\beqa
\langle\langle {\cal L}F| {\cal L}G\rangle\rangle&=& \langle\langle F| G\rangle\rangle\;\;\;  ({\cal L}\; {\rm unitary}),
\\
\langle\langle {\cal L}F| {\cal L} G\rangle\rangle&=& \langle\langle  F| G\rangle\rangle^*\;\;\;  ({\cal L}\; {\rm antiunitary}).
\eeqa
%
%
These transformations preserve the inner product (transition probability) of general (density operator) states, so they represent a mild generalization of Wigner's
formulation of symmetries \cite{Simon2018}.
Moreover they satisfy ${\cal L}^\dagger={\cal L}$, where the adjoints are defined differently for linear or antilinear superoperators,
\beqa
\langle\langle F| {\cal L}^\dagger G\rangle\rangle&=& \langle\langle {\cal L} F| G\rangle\rangle\;\;\;  ({\cal L}\; {\rm unitary}),
\\
\langle\langle F| {\cal L}^\dagger G\rangle\rangle&=& \langle\langle {\cal L} F| G\rangle\rangle^*\;  ({\cal L}\; {\rm antiunitary}).
\eeqa
The set $\{1, \cal{I,T,C,CT,TI,IC,CTI}\}$ forms the elementary abelian group $E8$ \cite{rose2009course}.
This is a homocyclic group, namely, the direct product of isomorphic
cyclic groups of order 2 with generators $\cal{C,T,I}$. We may, similarly to Eq. (\ref{defs}), define primmed superoperators in momentum representation, e.g. ${\cal T'} H =\int\!\!\int |p\ra \la p'|H|p\ra\la p'| dp dp'$. They also form the E8 group
$\{1, \cal{I',T',C',C'T',T'I',I'C',C'T'I'}\}$. Only for the subgroup $\{1, \cal{I,CT,CTI}\}$ the superoperators have the same representation-independent form in terms of complex conjugation, transposition and inversion.
%
%
%

%
\section{S-matrix pole structure}
\label{sec:SPoles}
%
%
%
The (on shell) $\sf{S}(p)$ matrix for $H$ is defined on the real positive momentum axis in terms of transmission and reflection amplitudes for right
and left incidence,
\begin{equation}
\sf{S}=\left(\begin{array}{cc}
T^l(p)&R^r(p)
\\
R^l(p)&T^r(p)
\end{array}\right).
\end{equation}
There is a companion matrix $\widehat{\sf{S}}$ with hatted amplitudes corresponding to scattering by $H^{\dagger}$. See Appendix \ref{sec:ScattFormalism} and \cite{Muga2004} for details.
For negative $p$ the matrix elements give the amplitudes of scattering states with a pure outgoing plane wave towards the right or the left.
Moreover we assume, as it is customary,
that the amplitudes may be continued analytically beyond the real axis.
The existence of a continuation on a complex plane domain depends on decay properties of the potentials and may
be checked in each case.
The analytical continuation is indeed possible for the model potentials of the following section.

The eigenvalues of $\sf{S}$ can be calculated from the transmission and reflection amplitudes as
\begin{equation}
S_j=\frac{(T^l+T^r)+(-1)^j[(T^l-T^r)^2+4R^lR^r]^{1/2}}{2}
\label{Sform}
\end{equation}
for $j=0,1$, and of course there is a similar expression for $\widehat{S}_j$ with hatted amplitudes.
They are related by \cite{Muga2004},
\beq\label{pam1}
S_j(p)=\wh{S}_j^*(-p^*)\,,
\eeq
and
\beq\label{pam2}
\wh{S}_j^*(p^*)S_j(p)=1\,.
\eeq
Combining Eqs. (\ref{pam1}) and (\ref{pam2}) gives
\beq\label{pam3}
S_j(p)=S^{-1}_j(-p)\,.
\eeq
%

When the following relations are fulfilled,
\beqa
T^{r,l}(p)&=&\widehat{T}^{r,l}(p)\; {\rm or}\; T^{r,l}(p)=\widehat{T}^{l,r}(p),
\label{ts}
\\
R^{r,l}(p)&=&\widehat{R}^{r,l}(p)\; {\rm or}\; R^{r,l}(p)=\widehat{R}^{l,r}(p),
\label{rs}
\eeqa
then
\beq
S_j(p)=\widehat{S}_j(p).
\eeq
Thus, from Eqs. (\ref{pam1}) and (\ref{pam2}), the poles and zeroes are found with the same symmetric pattern with
respect to the imaginary axis as for the Hermitian case (see Fig. \ref{fig:DiagramPoles}), the only difference being the possibility
of finding pairs of symmetrical poles in the upper complex plane when $H\neq H^\dagger$. They represent normalizable ``bound states
with complex energies''.
When they  are not present, the discrete spectrum becomes purely real.

One may check from Table I that Eqs. (\ref{ts}) and (\ref{rs}) are fulfilled
for symmetries  II (Hermiticity), VII (PT-symmetry), IV (parity pseudohermiticity),
and V (time-reversal invariance).  For local potentials these last two symmetries coalesce with the first two well-known
cases \cite{Ruschhaupt2017}, namely,
IV becomes equivalent to PT-symmetry, and V becomes equivalent to Hermiticity. For non-local potentials, though, these symmetries
correspond to genuinely distinct properties. In the following section we shall demonstrate this fact with potentials that are
purely parity-pseudohermitian (and not PT-symmetrical), and time-reversal invariant but not Hermitian.

%
%
%
%
%
%
%
\section{Separable Potentials}
\label{sep_pot_sec}
Separable potentials are quite useful models as a solvable approximation to realistic ones.
Often they lead to explicit expressions
for wave functions or scattering amplitudes, so they are used to test concepts and new methods.
They are also instrumental in learning about different dynamical phenomena (for example transient effects, short-time and long-time behavior, or anomalous decay laws)  and their relation to complex-plane singularities
 \cite{Muga1990,Muga1996,Muga1996b,Muga1998a}. Their simplest version takes the form
$|\chi\ra V_0\la\chi|$.   In particular, with a complex $V_0$,
 they have been used to examine anomalous (negative) time delays caused by  crossing of zeroes of the $S$-matrix eigenvalues or $S$-matrix elements across the momentum real axis \cite{Muga1998b}.

In this work we consider the simple structure
$V=V_0 \ketbra{\phi}{\chi}$, with $V_0$ real. The main aim is to set explicit models to demonstrate the formal results of the previous section, in particular the
symmetrical configuration of poles with respect to the imaginary axis in the complex momentum plane for certain Hamiltonian symmetries.
In passing we shall also note some interesting phenomena that may be studied in more detail elsewhere, such as pole collisions, crossings of
the real axis, or diodic (Maxwell demon) behavior with asymmetrical transmission for right/left incidence.

From the stationary Schr\"{o}dinger equation $H \ket{\psi} = E \ket{\psi}$, the eigenvalues may be found by solving
\begin{eqnarray}
Q_{0}(E)V_{0} = 1,
\label{roots}
\end{eqnarray}
where $Q_{0}(E)=\bra{\chi}(E-H_0)^{-1}\ket{\phi}$ and $H_{0}=p^{2}/(2m)$.

Moreover, for a separable potential $V_0 \ketbra{\phi}{\chi}$, the transition operator $T_{op}$ can be written (see Appendix \ref{app1}) as
\begin{equation}
T_{op}=\frac{V_{0}}{1-V_{0} Q_{0}(E)} \ketbra{\phi}{\chi}.
\end{equation}
Since all scattering amplitudes in $S$ are simply related to matrix elements of $T_{op}$ in momentum representation, see
Eq. (\ref{art}), solutions to Eq. (\ref{roots}) provide their core singularities (independent of the representation \cite{Muga1996}).

Once $Q_{0}(E)$  is calculated, the transmission and reflection amplitudes can be found from \eqref{art}
using the momentum representation of $\ket{\phi}$ and $\ket{\chi}$.
\subsection{Time-reversal symmetric potential}
\begin{figure}[ht]
    \includegraphics[width=0.75\linewidth]{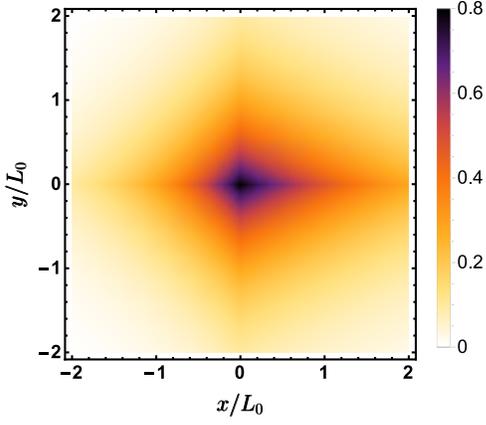}
    \caption{(Color online) $\la x|V|y\ra L_0 / V_0$ for the time-reversal symmetric potential \eqref{TRpot} with parameters $a = p_0$, $b = 0.5\, p_0$, and $V_0>0$.}
    \label{fig:VSymPotentialPlot}
\end{figure}

\begin{figure}[h]
    \includegraphics[width=1.0\linewidth]{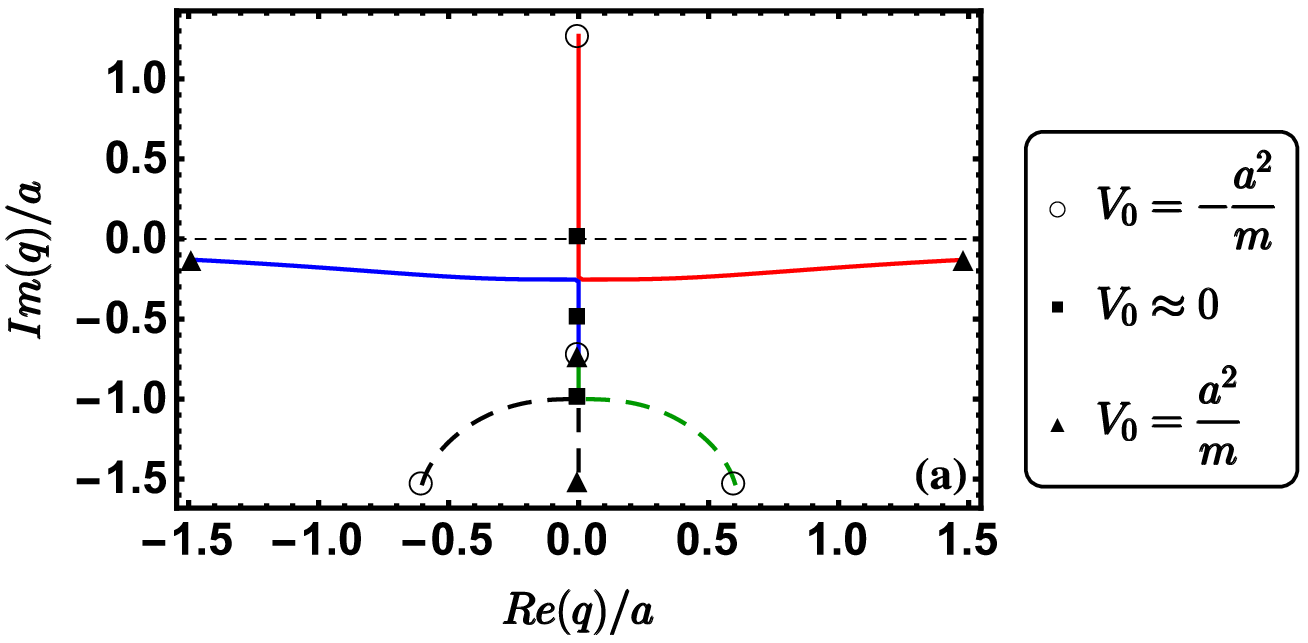}
    \includegraphics[width=1.0\linewidth]{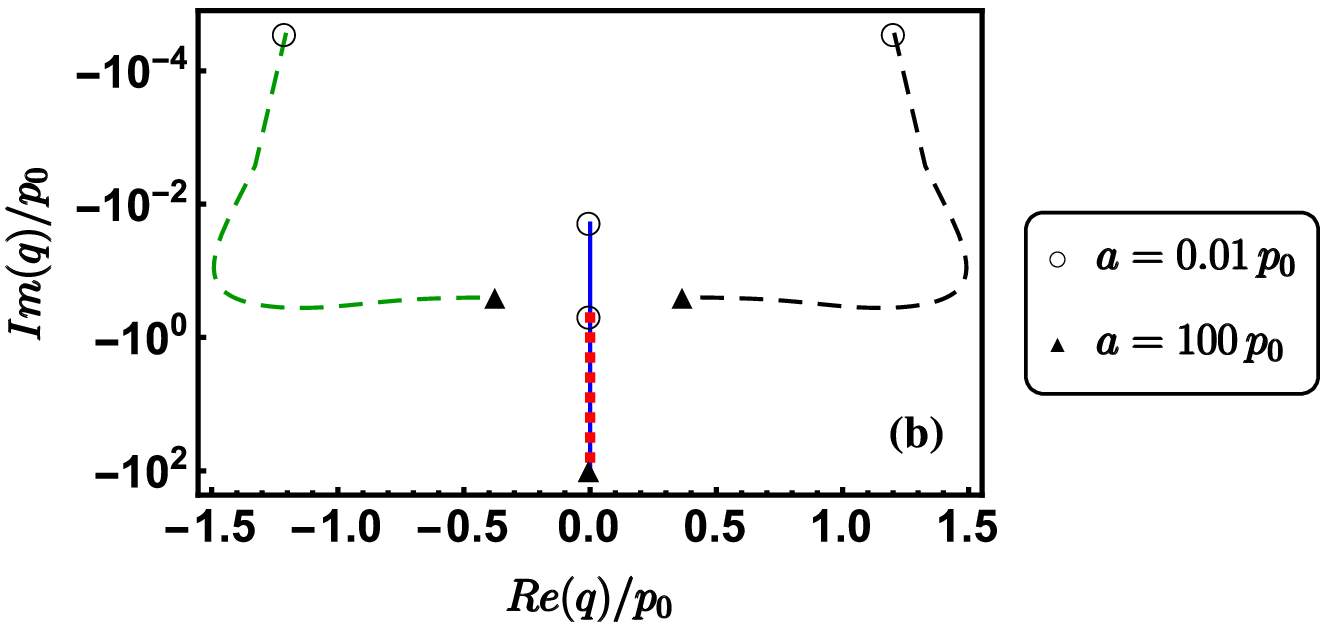}
    \includegraphics[width=1.0\linewidth]{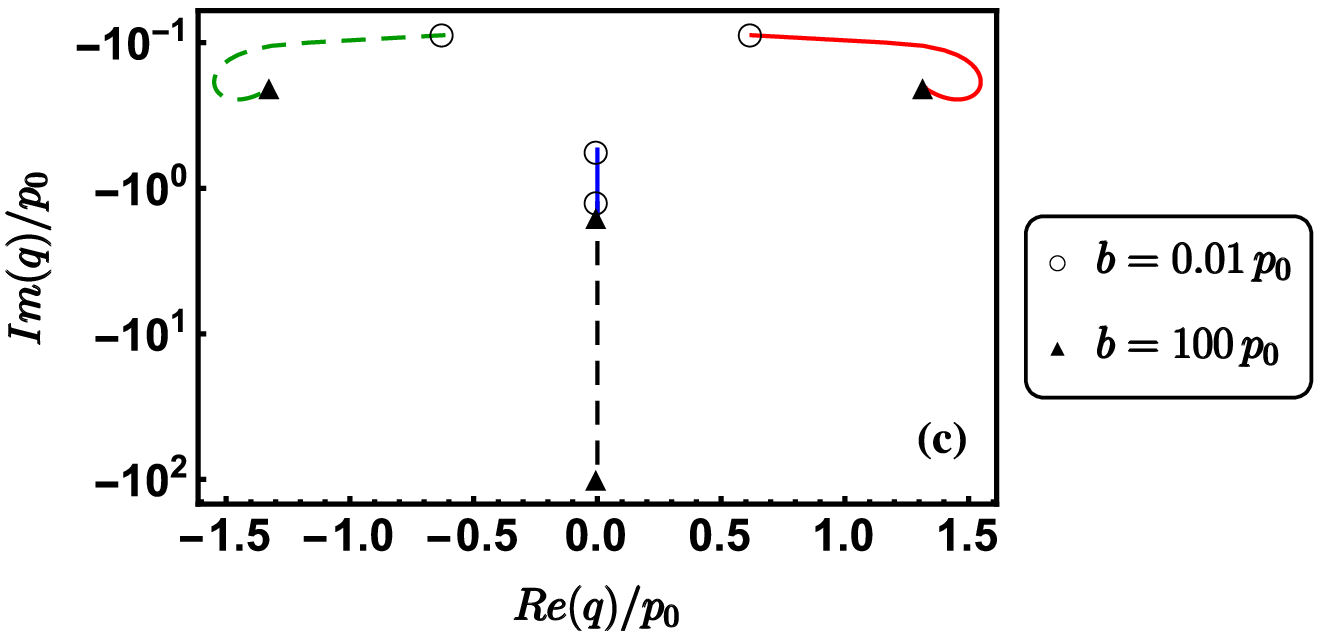}
    \caption{(Color online) Poles and pole trajectories of time-reversal symmetric potential \eqref{TRpot} for (a) varying $V_0$ with $a=2 b$; (b) varying $a$ with $b=0.5\, p_0$, $V_0>0$; and (c) varying $b$ with $a=p_0$,
    $V_0>0$. At pole collisions we connect each of the incoming trajectories with a different emerging trajectory but the choice of outgoing branch  is arbitrary since the two colliding poles lose their identity.}
    \label{fig:VSymEigenvals}
\end{figure}

We start with an example of a separable potential which only satisfies symmetry V (apart from the trivial symmetry I). The normalised vector $\ket{\chi}$, is given in position and momentum representation as
\begin{eqnarray}
\braket{x}{\chi}&=&\sqrt{\frac{a}{\hbar}} e^{-a \fabs{x}/\hbar}, \nonumber \\
\braket{p}{\chi}&=& \sqrt{\frac{2 a^3}{\pi}} \frac{1}{p^2+a^2}.
\end{eqnarray}
We choose $\ket{\phi}$ similarly as
\begin{eqnarray}
\braket{x}{\phi}=&\sqrt{\frac{2ab}{\hbar (a+b)}} \begin{cases}
 e^{-b x/\hbar} &x>0,\\ e^{a x/\hbar}  &x<0,
   \end{cases}\nonumber \\
\braket{p}{\phi}=& \sqrt{\frac{ab}{\pi (a+b)}}\frac{a+b}{(p+i a)(p-i b)},
\end{eqnarray}
where $a,b>0$. In coordinate representation the potential is given as
\begin{eqnarray}
\la x|V|y\ra = V_{0} \sqrt{\frac{2 b a^2}{\hbar^2 (a+b)}} \begin{cases}
 e^{-(a \fabs{y}+b x)/\hbar} \, &x>0,\\ e^{a (x-\fabs{y})/\hbar} \,  &x<0,
   \end{cases} \label{TRpot}
\end{eqnarray}
which can be seen in Fig. \ref{fig:VSymPotentialPlot}. Clearly the potential is always even in $y$ and in the case where $a=b$, is also even in $x$. For $a=b$, the potential will satisfy parity symmetry (III) and hence also PT symmetry (VII). In this case, there is no asymmetric transmission or reflection.

We define first a complex momentum $q=\sqrt{2 m E}$ (for complex $E$) with positive imaginary part.
To calculate $Q_{0}(q)$ explicitly we use a closure relation in momentum representation, and
complex contour integration around the poles at $ia$, $q$ and $ib$.
The result is then analytically continued to the whole $q$-plane,
\beqa
&&Q_{0}(q)/m=
\nonumber\\
&&-\frac{i \sqrt{2b} \left[2 a (a+b)^2-q^2 (3 a+b)-i q (2 a+b) (3 a+b)\right]}{q (a+b)^{3/2} (a-i q)^2 (b-i q)},
\nonumber\\
\label{eq:ResolvantVSymm}
\eeqa
with which we may calculate the transmission and reflection amplitudes.
The four roots of Eq. (\ref{roots}) are the core poles.

Using $m$, $V_0$ and $\hbar$ we define the length and momentum scales $L_0 = \hbar/\sqrt{mV_0}$ and $p_0 = \sqrt{mV_0}$. In Fig. \ref{fig:VSymEigenvals}(a), we can see the trajectory of the $S$-matrix core poles (zeros
of $1-V_0Q_0(q))$ for varying $V_0$. Notice a bound state for $V_0<0$ and collisions of the eigenvalue pairs around $V_0 = 0$. In Figs. \ref{fig:VSymEigenvals}(b) and \ref{fig:VSymEigenvals}(c), where $V_0$ is positive and $a$ or $b$ are varied,
there are two virtual states and one resonance/anti-resonance pair. In all cases the symmetry of the poles about the imaginary axis
which corresponds to real energies or complex-conjugate pairs of energies, is evident. For larger values of the $a$ or $b$ parameters
(not shown)
the pair collides so that all poles end up as virtual states.

\begin{figure}
    \begin{center}
    \includegraphics[width=1\linewidth]{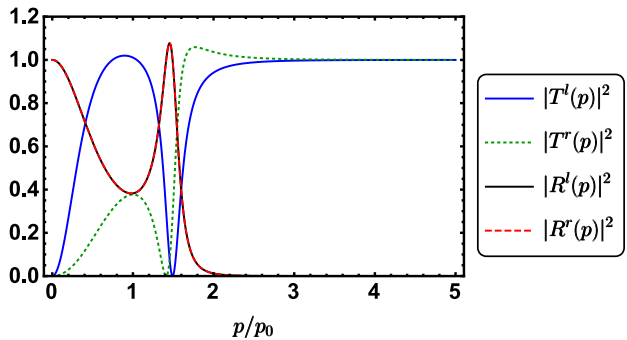}
    \end{center}
    \caption{(Color online) Transmission and reflection coefficients of the time-reversal symmetric potential \eqref{TRpot} with $a=p_0$, $b= 0.5\, p_0$ and $V_0>0$.}
    \label{fig:VSymScattAmplitudes}
\end{figure}

Figure \ref{fig:VSymScattAmplitudes} depicts the associated transmission and reflection coefficients (square moduli of the amplitudes) as functions of the momentum $p$. $|R^l(p)|=|R^r(p)|$ for all $p$ due to symmetry V \cite{Ruschhaupt2017}.
The coefficients can be greater than one (in contrast to the Hermitian case).

\begin{figure}
    \begin{center}
    \includegraphics[width=1\linewidth]{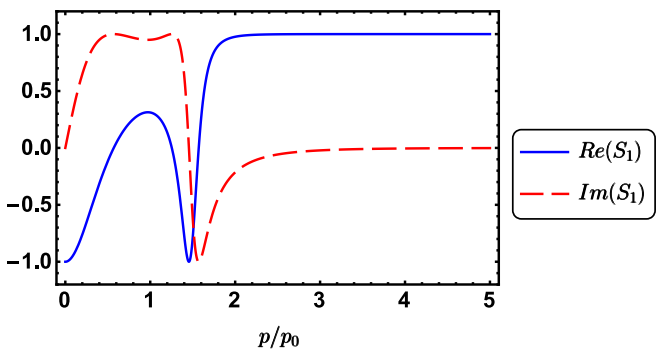}
    \end{center}
    \caption{(Color online) $S_1$ eigenvalue of the  $S$-matrix of the time-reversal symmetric potential \eqref{TRpot} with $a= p_0$, $b= 0.5\, p_0$ and $V_0>0$.}
    \label{fig:VSymSEigenvalue}
\end{figure}

In Fig. \ref{fig:VSymSEigenvalue}, the $S_1$ eigenvalue of the scattering matrix is shown. {From Eq. \eqref{pam3}, at $p=0$ the $S$-matrix eigenvalues must be either $\pm 1$. Since $S_2=1$ for all values of $p$ (See Appendix \ref{app2}), $S_1=-1$ at $p=0$.} Also, since there is full transmission for very high kinetic energies, $S_j \rightarrow 1$ as $p \rightarrow \infty$. Both of these limiting properties can be seen in Fig. \ref{fig:VSymSEigenvalue}. The crossover into the regime where the kinetic energy dominates can be seen for $E_p \gg V_0$, i.e. when $p/p_0 \gg \sqrt{2}$.
\subsection{Parity pseudohermitian potential}
\begin{figure}[h]
    \includegraphics[width=0.9\linewidth]{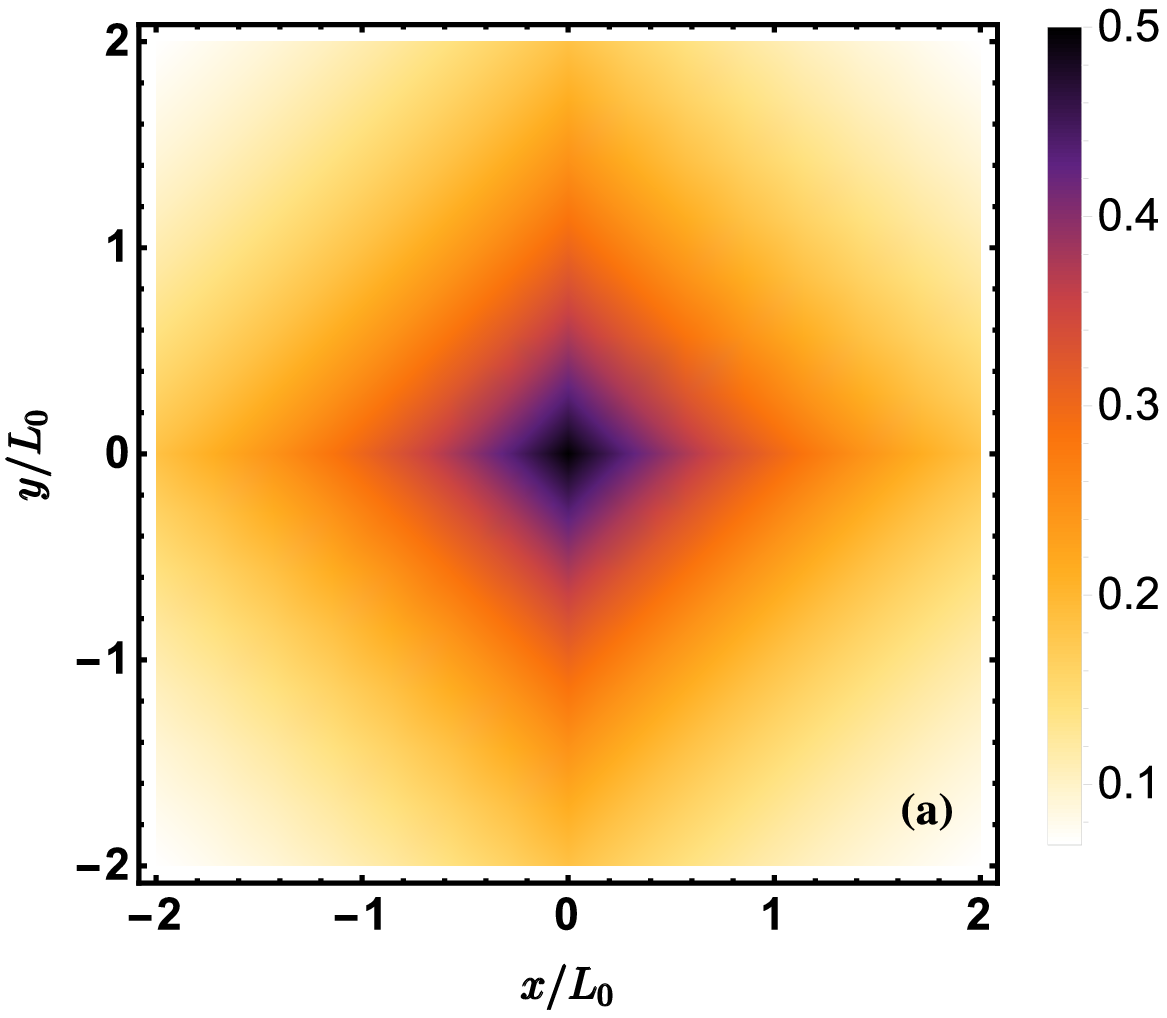}
    \includegraphics[width=0.9\linewidth]{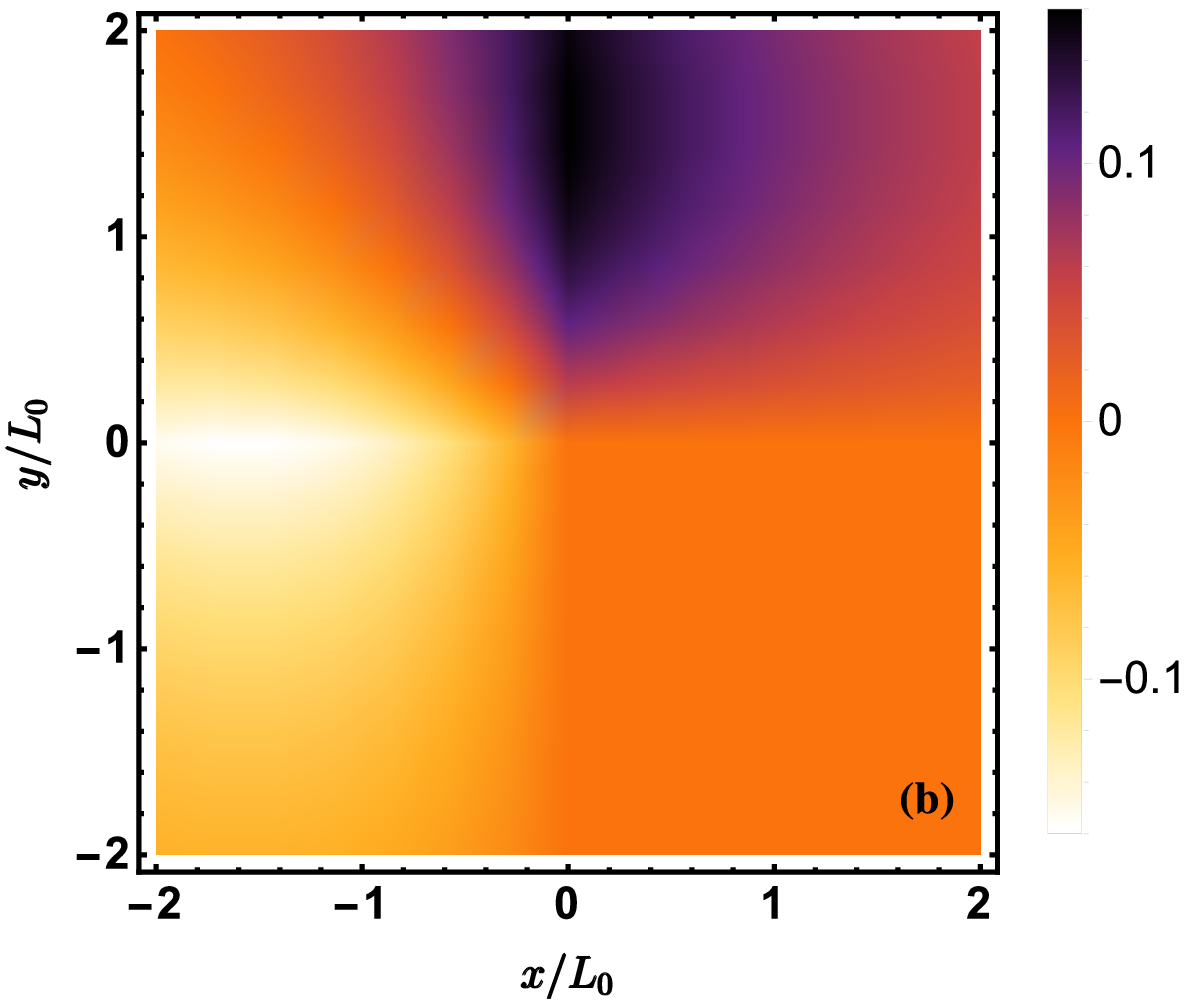}
    \caption{(Color online) The (a) real and (b) imaginary parts of $\la x|V|y\ra L_0 / V_0$ for the parity pseudohermitian potential \eqref{Ppot} with parameters $a= b =0.5\, p_0$, $V_0>0$.}
    \label{fig:IVSymPotentialPlot}
\end{figure}

As a second example we will consider  a separable potential which only fulfils symmetry IV. The normalised vector $\ket{\chi}$ in position and momentum representation is
\begin{eqnarray}
\braket{x}{\chi}=& \sqrt{\frac{a}{\hbar}} \begin{cases}
e^{-(a+ib)x/\hbar}  &x>0,\\ e^{a x/\hbar} &x<0,
   \end{cases} \nonumber \\
\braket{p}{\chi}=&  \sqrt{\frac{a}{2\pi}} \frac{2 a+ i b}{(p+ia)(p+b-i a)},
\end{eqnarray}
where $a>0$ and $b$ is real. We choose $\ket{\phi}$ as
\begin{eqnarray}
\braket{x}{\phi}=& \sqrt{\frac{a}{\hbar}} \begin{cases}
e^{-a x/\hbar} &x>0,\\ e^{(a+i b)x/\hbar} &x<0,
   \end{cases} \nonumber \\
\braket{p}{\phi}=& \sqrt{\frac{a}{2\pi}} \frac{2 a+i b}{(p-ia)(p-b+i a)} .
\end{eqnarray}
In coordinate representation the potential is
\begin{eqnarray}
\la x|V|y\ra=  \frac{aV_{0}}{\hbar} \begin{cases}
 e^{-\left[a (x+y) - i b y\right]/\hbar} \, ,&x>0,\,y>0\\
 e^{a(y-x)/\hbar} \,  ,&x>0, \,y<0\\
 e^{\left[a(x-y)+i b(x+y)\right]/\hbar} \,  ,&x<0,\,y>0\\
 e^{\left[a(x+y)+i b x\right]/\hbar} \,  ,&x<0,\,y<0
   \end{cases},
   \label{Ppot}
\end{eqnarray}
see Fig. \ref{fig:IVSymPotentialPlot}. The case $b=0$ implies that the potential is real and hence satisfies time-reversal symmetry (V) with equal reflection amplitudes (as in the previous case), and also symmetry VIII.

\begin{figure}[h]
    \includegraphics[width=1.0\linewidth]{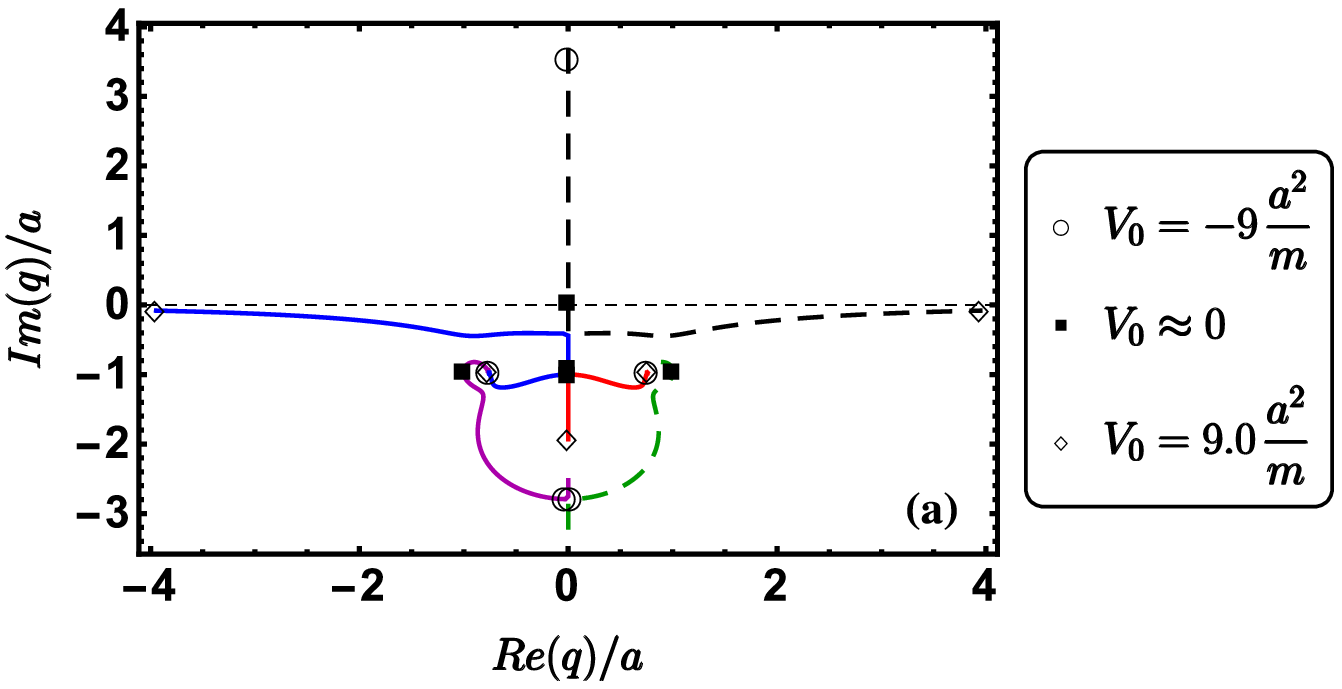}
    \includegraphics[width=1.0\linewidth]{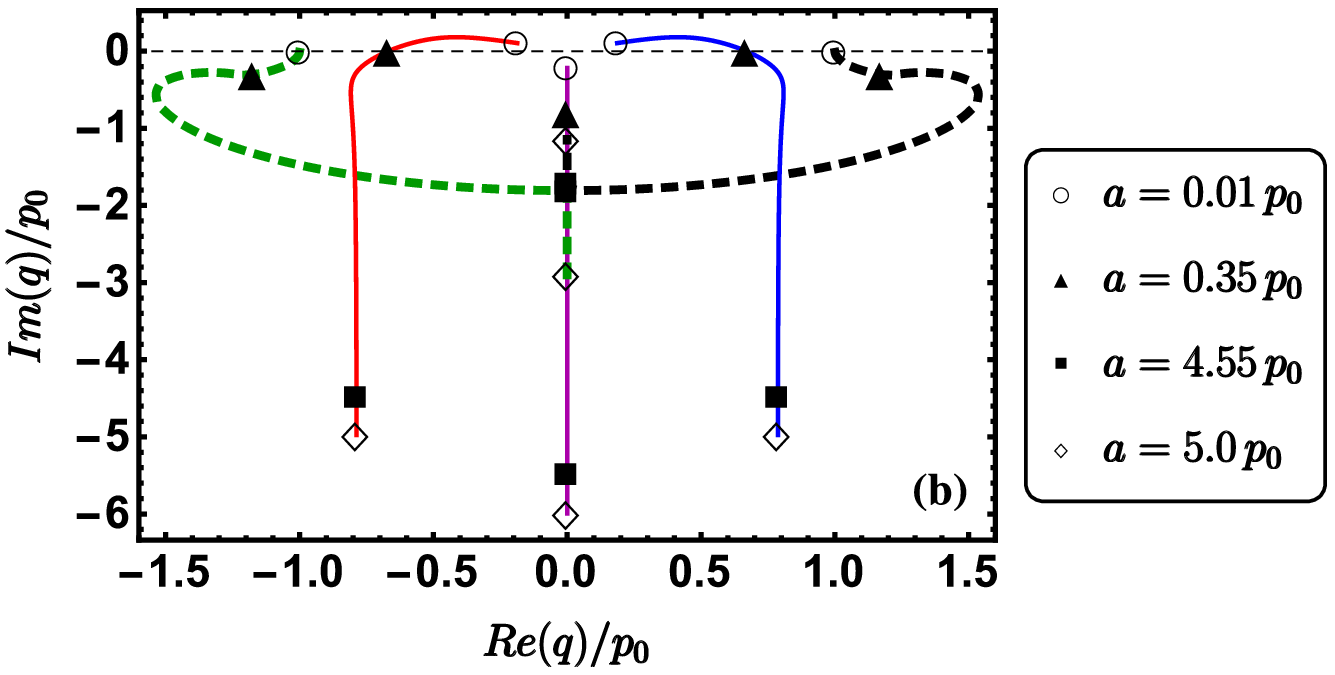}
    \includegraphics[width=1.0\linewidth]{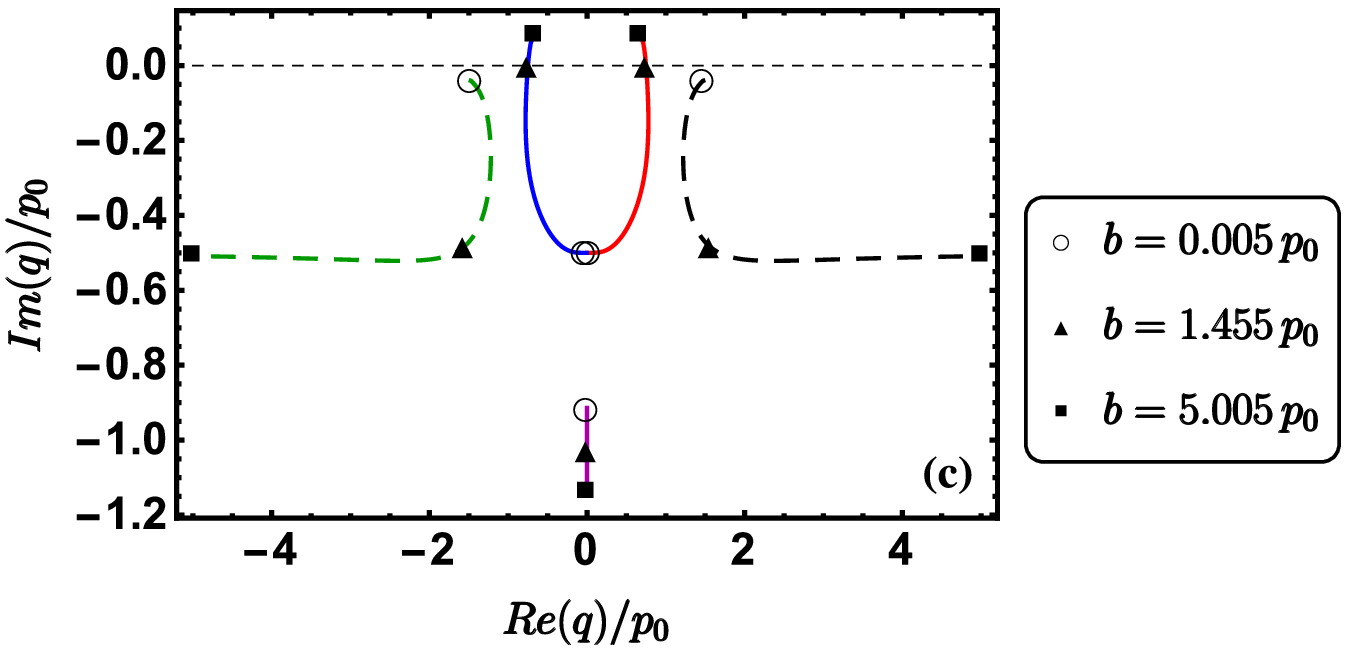}
    \caption{(Color online) Poles and pole trajectories for the parity pseudohermitian potential \eqref{Ppot} (a) varying $V_0$ with $a=b$; (b) varying $a$ with $b=p_0$, $V_0>0$; (c) varying $b$ with $a=0.5\, p_0$, $V_0>0$.}
    \label{fig:IVSymEigenvals}
\end{figure}

By calculating $Q_{0}$ again explicitly using complex contour integration around the poles at $-q$, $-b-i a$ and $b-i a$, we get that
\beqa
&&Q_{0}(q)/m
\nonumber\\
&&=\frac{8 a^2 q^3-4 a^2 q \left(10 a^2+b^2\right)-i a \left(4 a^2+b^2\right)^2+32 i a^3 q^2}{q \left(4 a^2+b^2\right) (a-i q)^2 \left[b^2+(a-i q)^2\right]}.
\nonumber\\
\eeqa
Equation \eqref{roots} has five roots in this case constituting core poles of the $S$ matrix elements.

Figure \ref{fig:IVSymEigenvals} depicts the trajectories of these poles for varying $a$, $b$ or $V_0$. As for the previous potential, the poles are symmetric with respect to the imaginary axis. In Fig. \ref{fig:IVSymEigenvals}(a) there is a single bound state for $V_0<0$ while for positive values there are a resonance/antiresonance pair and a pair of virtual states. There are collisions of eigenvalues for values of $V_0$ close to 0. In Fig. \ref{fig:IVSymEigenvals}(b)  two complex-conjugate (bound) eigenvalues cross the real axis and become a resonance/antiresonance pair. At the exact point where the eigenvalues are on the real axis, the scattering amplitudes diverge, however the eigenvalues of the $S$ matrix do not, since divergences of the left and right amplitudes cancel each other. For $a  \approx 4.55$ $p_0$ a resonance/antiresonance pair collides and becomes a pair of virtual states. In Fig. \ref{fig:IVSymEigenvals}(c) another crossing of the real axis takes place, but in this case when decreasing $b$.

\begin{figure}[t]
\begin{center}
\includegraphics[width=\linewidth]{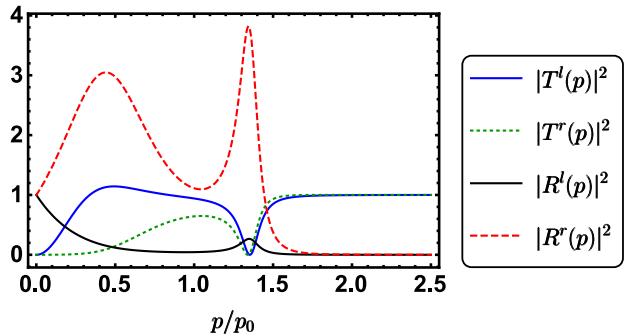}
\end{center}
\caption{(Color online) Transmission and reflection coefficients for $a=b=0.5\, p_0$ and $V_0>0$.}
\label{fig:T_R_fig2}
\end{figure}

Figure \ref{fig:T_R_fig2} depicts the associated transmission and reflection coefficients as functions of the momentum $p$. The eigenvalues are not always equal since parity pseudohermicity does not imply any strict restriction to them \cite{Ruschhaupt2017}. For large  momenta, i.e. $p \gg \sqrt{2} p_0$, the potential is transparent giving $T^l,T^r \approx 1$. For $p\approx 1.5$ $p_0$ the right incidence transmission has a pronounced peak. Comparing with \ref{fig:IVSymEigenvals}(c), we notice that the values of the potential parameters and the momentum are close to the ones for which the real axis crossing takes place. Around $p = 0.6$ $p_0$ the potential acts as an asymmetric transmitter
\cite{Ruschhaupt2017}.

\begin{figure}[t]
\begin{center}
\includegraphics[width=\linewidth]{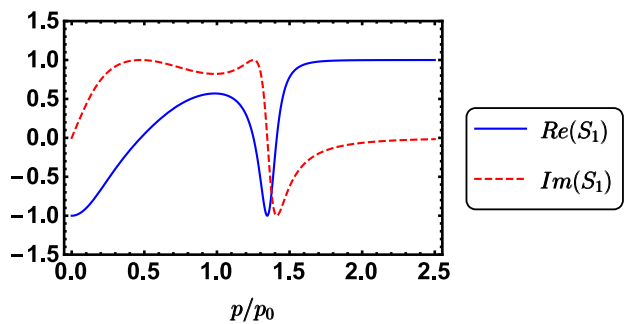}
\end{center}
\caption{(Color online) $S$-matrix eigenvalue $S_1$ for $a=p_0$ and $b=0.5\, p_0$ and $V_0>0$.}
\label{fig:IVSeigs}
\end{figure}

Figure \ref{fig:IVSeigs} shows the real and imaginary part of the first eigenvalue $S_1$ of the scattering matrix. Since the potential is separable $S_2 = 1$ (see Appendix \ref{app2}). For $p$ going to infinity there is no scattering and the scattering matrix is the identity.

%

\section{Conclusion}
\label{sec:Conclusions}

In this paper we have studied some aspects of the scattering of a structureless particle in one dimension by
generally non-local and non-Hermitian potentials.
Conditions that were found for discrete Hamiltonians to imply conjugate pairs of discrete eigenenergies
(pseudohermiticity with respect to a linear operator or commutativity of $H$ with an antilinear operator \cite{Mosta2002a,Mosta2002b,Mosta2002c})
can in fact be extended to scattering Hamiltonians in the continuum, implying symmetry relations not just for bound-state eigenvalues
but also for complex
poles of the $S$-matrix. Specifically the poles of $S$ matrix eigenvalues
are symmetrically located with respect to the imaginary axis, also in the lower momentum plane, so that resonances and antiresonance
energies are conjugate pairs as well.
In  terms of the eight possible Hamiltonian symmetries associated with Klein's group of $A$ operators (unity, parity, time reversal and PT)
and their commutation or pseudohermiticity with $H$,
the symmetrical disposition of the poles applies to four of them, which includes hermiticity and PT-symmetry. Potential models
and pole motions are provided for the
two other non trivial symmetries: time-reversal symmetry and parity pseudohermiticity.

For future work we plan to consider more complicated systems including internal states, as well as physical realizations of the different
symmetries.

\acknowledgments{
This work was supported by the Basque Country Government (Grant No. IT986-16),  and MINECO/FEDER,UE (Grant No. FIS2015-67161- P). M.A. Sim\'on acknowledges support by the Basque Government predoctoral program (Grant no. PRE-2017-2-0051).
AM wishes to acknowledge the support provided by the Turkish Academy of Sciences (T\"urkiye Bilimler Akademisi) through its membership grant.}

\appendix

\section{Review of scattering theory formalism \label{sec:ScattFormalism}}
A detailed overview of scattering theory can be found in \cite{taylor2006scattering} and its extension to NH systems in \cite{Muga2004}. Scattering theory describes the interaction of an incoming wave packet with a localized potential. In general, the spectrum of scattering Hamiltonians has both a discrete part and a continuum with real, positive energies.
The eigenstates of the continuous spectrum are constructed by the action on plane waves of the M\"oller operators
$\ket{p^\pm}=\Omega_\pm \ket{p}$ and $\ket{\widehat{p}^\pm}=\widehat{\Omega}_\pm \ket{p}$,
where
%
\begin{eqnarray}
    \Omega_+ &=& \lim_{t \to -\infty}e^{i H t / \hbar}e^{-i H_0 t/ \hbar},\nonumber\\
    \Omega_- &=& \lim_{t \to \infty}e^{i H\da t/ \hbar}e^{-i H_0 t/ \hbar},\nonumber\\
    \widehat{\Omega}_+ &=& \lim_{t \to -\infty}e^{i H\da t/ \hbar}e^{-i H_0 t/ \hbar},\nonumber\\
    \widehat{\Omega}_- &=& \lim_{t \to \infty}e^{i H t/ \hbar}e^{-i H_0 t/ \hbar},
\end{eqnarray}
and a regularization of the limit is implied, see e.g. \cite{Muga2004}.
The M\"oller operators satisfy the isometry relation $\widehat{\Omega}_{\pm}\da\Omega_{\pm} = 1$ and the interwining relations $H \Omega_+ = \Omega_+ H_0$ and $H\da \Omega_- = \Omega_- H_0$.
By using the intertwining relations, it is easy to see that $\ket{p^+}$ and $\ket{\widehat{p}^-}$ are right eigenvectors of $H$ while $\ket{\widehat{p}^+}$ and $\ket{p^-}$ are left eigenvectors of $H$, all with positive energy $E_p = p^2/2m$. In the following we will assume that the Hamiltonian admits a basis of biorthonormal
right/left eigenstates $\left\{ \ket{\psi_n} , \ket{\phi_a} \right\}$ with energies $E_n$ satisfying $\braket{\phi_n}{\psi_m} = \delta_{n,m}$ for the discrete part. The stationary scattering states are also biorthonormal, i.e. $\braket{\widehat{p}^+}{q^+} = \braket{\widehat{p}^-}{q^-} = \delta (p-q)$ and together with the eigenstates of the discrete spectrum they give the resolution of the identity
\begin{eqnarray}
    1 &=& \sum_{n}\ketbra{\psi_n}{\phi_n} + \int_{-\infty}^{\infty} dp\, \ketbra{p^+}{\widehat{p}^+}
    \nonumber\\
    &=& \sum_{n}\ketbra{\psi_n}{\phi_n} + \int_{-\infty}^{\infty} dp\, \ketbra{\widehat{p}^-}{p^-}.
    \label{eq:identity}
\end{eqnarray}
Let us note that there is no degeneracy in the discrete spectrum of one-dimensional systems, whereas the continuum is doubly degenerate,
e.g. with continuum eigenfunctions incident from the right or the left.  We shall explicitly make use of this property in what follows. Using the resolution of the identity in terms of discrete eigenstates and the stationary scattering states, the Hamiltonian can be expanded as
\begin{equation}
    H = \sum_{n}E_n\ketbra{\psi_n}{\phi_n} + \frac{1}{2m}\int_{-\infty}^{\infty} dp\,p^2 \ketbra{p^+}{\widehat{p}^+}.
    \label{eq:H_Expansion_Continuous}
\end{equation}
We call the first and the second terms of \eqref{eq:H_Expansion_Continuous} the discrete, $H_d$, and continuous, $H_c$, parts of the Hamiltonian respectively. A central object is the scattering operator (or matrix), $S \equiv \Omega_-\da \Omega_+$ for scattering processes by $H$ and $\widehat{S} \equiv \widehat{\Omega}_-\da \widehat{\Omega}_+$ for $H\da$. Unhated quantities refer to scattering by $H$, while hated ( $\widehat{\;}$ ) quantities refer to scattering by its Hermitian conjugate $H\da$. The scattering operator gives the probability of an incident state $\ket{\psi_{in}}$ to be scattered (by $H$ or $H \da$) into a state $\ket{\psi_{out}}$ as $\left|\bra{\psi_{out}} S \ket{\psi_{in}}\right|^2$ or $\left|\bra{\psi_{out}} \widehat{S} \ket{\psi_{in}}\right|^2$. Although the scattering operator is not unitary for NH Hamiltonians, $S$ and $\widehat{S}$ obey the relation $\widehat{S}\da S = 1$ which collapses to the unitarity condition ($S = \widehat{S}$) if $H = H\da$. If the Hamiltonian is symmetric or pseudo-hermitian with respect to a linear/antilinear operator $A$, the M\"oller and scattering operators transform according to the intertwining relations in Table \ref{tab:MollerOperatorSyms}. The intertwining relations of the M\"oller operators give the transformation rules for scattering states under $A$ and provide interesting relations between the different transmission/reflection coefficients.

\begin{table}
  \centering
  \begin{tabular}{|c|c|c|}
  \hline
   & \textbf{$A$ linear} & \textbf{$A$ antilinear}

  \\
  \hline

  $A H = H A$
  &
  $
  \begin{array}{ccc}
    &&
    \\
    A \Omega_{\pm}&=&\Omega_{\pm} A
    \\
    A S&=&S A
    \\
    &&
  \end{array}
  $
  &
  $
  \begin{array}{ccc}
    &&
    \\
    A \Omega_{\pm}&=&\widehat{\Omega}_{\mp} A
    \\
    A S&=&\widehat{S}^\dagger A
    \\
    &&
  \end{array}
  $
  \\
  \hline

  $A H = H^\dagger A$
  &
  $
  \begin{array}{ccc}
    &&
    \\
    A \Omega_{\pm}&=&\widehat{\Omega}_{\pm} A
    \\
    A S&=&\widehat{S} A
    \\
    &&
  \end{array}
  $
  &
  $
  \begin{array}{ccc}
    &&
    \\
    A \Omega_{\pm}&=&\Omega_{\mp} A
    \\
    A S&=&S^\dagger A
    \\
    &&
  \end{array}
  $
  \\
  \hline
  \end{tabular}
  \caption{Transformation rules of the M\"oller and scattering operators under symmetries/pseudo-symmetries with linear or antilinear operators.}
   \label{tab:MollerOperatorSyms}
\end{table}

Also relevant to scattering theory is the transition operator, which is defined as
\begin{eqnarray}
T_{op}(E)=V+VG(E)V
\end{eqnarray}
where $G(E)=(E-H)^{-1}$ is Green's operator. The transition operator satisfies $T^{\dagger}_{op}(z)=\wh{T}_{op}(z^*)$ and its matrix elements in momentum representation are related to the the scattering operator by
\begin{equation}
    \bra{p} S \ket{p'} = \delta (p-p') -2 i \pi \delta (E_p-E_{p'}) \bra{p}T_{op}(+)\ket{p'},
\end{equation}
where  $T_{op}(\pm)|p'\ra=\lim_{\epsilon\to0^+} T_{op}(E_p\pm i\epsilon)|p'\ra$. This operator can then be used to define the scattering amplitudes
for real $p$ as
\beqa\label{art}
R^l(p)&=&-\frac{2\pi im}{p} \la -p|T_{op}({\rm sign}(p))|p\ra\,, \nn\\
T^l(p)&=&1-\frac{2\pi im}{p} \la p|T_{op}({\rm sign}(p))|p\ra\,, \nn\\
R^r(p)&=&-\frac{2\pi im}{p} \la p|T_{op}({\rm sign}(p))|-p\ra\,, \nn\\
T^r(p)&=&1-\frac{2\pi im}{p} \la -p|T_{op}({\rm sign}(p))|-p\ra,
\eeqa
where $R^{l,r}$ is the left/right reflection amplitude and $T^{l,r}$ is the left/right transmission amplitude. We assume that the amplitudes admit analytic continuations. The generalized unitarity relation of the scattering operators give the following set of equations for the amplitudes
\begin{eqnarray}
    \widehat T^l(p) T^{l*}(p) + \widehat R^l(p) R^{l*}(p) &=& 1,
    \nonumber\\
    \widehat T^r(p) T^{r*}(p) + \widehat R^r(p) R^{r*}(p) &=& 1,
    \nonumber\\
    \hat T^{l*}(p) R^r(p) + T^r(p) \widehat R^{l*}(p) &=& 0,
    \nonumber\\
    T^l(p) \widehat R^{r*}(p) + \widehat T^{r*}(p) R^l(p) &=& 0,
    \label{gurAmplitudes}
\end{eqnarray}
where $p$ is taken to be real and nonnegative.

\section{Properties of separable potentials}
\subsection{Transition operator}
\label{app1}
 For a separable potential $V=V_0 \ketbra{\phi}{\chi}$, the transition operator becomes
\begin{eqnarray}
T_{op}=\alpha \ketbra{\phi}{\chi}
\end{eqnarray}
where $\alpha=V_{0}+V_{0}^{2}\bra{\chi}G(E)\ket{\phi}$. Then using the Lippmann-Schwinger equation we get that
\begin{eqnarray}
T_{op}(E)&=&V+VG_{0}(E)T_{op}(E)
\nonumber \\
&=& \left[V_{0}+\alpha V_{0} \bra{\chi}G_{0}(E)\ket{\phi}\right]\ketbra{\phi}{\chi}
\end{eqnarray}
where $G_{0}(E)=(E-H_{0})^{-1}$ is the Green's operator for free motion. Solving for $\alpha$ now gives
\begin{eqnarray}
\alpha =\frac{V_{0}}{1-V_{0} \bra{\chi}G_{0}(E)\ket{\phi}}
		   =\frac{V_{0}}{1-V_{0} Q_{0}(E)}.
\end{eqnarray}
%
%
%
%
%
\subsection{$S$-matrix eigenvalues}
\label{app2}
The eigenvalues for the S-matrix are given by Eq. \eqref{Sform} in terms of the reflection and transmission amplitudes. For a separable potential, using Eq. \eqref{art}, we can simplify the transmission and reflection coefficients as
\begin{eqnarray}
T^{l}&=&1-\frac{2 \pi i m}{p} \alpha \phi(p) \chi^{*}(p),
\nonumber \\
T^{r}&=&1-\frac{2 \pi i m}{p} \alpha \phi(-p) \chi^{*}(-p),
\nonumber \\
R^{l}&=&-\frac{2 \pi i m}{p} \alpha \phi(-p) \chi^{*}(p),
\nonumber \\
R^{r}&=&-\frac{2 \pi i m}{p} \alpha \phi(p) \chi^{*}(-p).
\nonumber \\
\end{eqnarray}
If we now define
\begin{equation}
\Gamma=\frac{2 \pi i m}{p} \alpha \left[\phi(p) \chi^{*}(p)+\phi(-p) \chi^{*}(-p)\right],
\end{equation}
we can write the eigenvalues as simply
\begin{eqnarray}
S_{j}&=&1-\frac{\Gamma-(-1)^{j}\Gamma}{2}.
\end{eqnarray}
Note that $S_2=1$ for all $p$. Clearly the following relation must also always hold for the reflection and transmission amplitudes,
\begin{equation}
T^l + T^r - T^l T^r + R^l R^r = 1.
\end{equation}
\subsection{Bound states}
\label{app3}
Note that a separable potential  can only have at most one bound state $\ket{\psi_{E}}$.
In momentum representation,
\begin{eqnarray}
\braket{p}{\psi_{E}}&=&\bra{p}\frac{V_{0}}{E-H_{0}}\ket{\phi}\braket{\chi}{\psi_{E}}
\nonumber \\
&=&\frac{M}{p^{2}-q_{B}^{2}}\braket{p}{\phi},
\end{eqnarray}
where $M=-2 m V_{0} \braket{\chi}{\psi_{E}}$ and $q_{B}^2=2 m E<0$. Suppose there is a second bound state $\ket{\psi_{E'}}$,
with corresponding quantities $M'$ and $q_{B'}^2$. Then,
\begin{eqnarray}
\braket{\psi_{E'}}{\psi_{E}}&=&M M' \int_{-\infty}^\infty dp \fabsq{\braket{p}{\phi}} \frac{1}{p^{2}-q_{B}^{2}}\frac{1}{p^{2}-q_{B'}^{2}}. \nonumber \\		\end{eqnarray}
Since $MM'\ne 0$ and the integral is positive the overlap cannot be zero
so there cannot be two bound states.
%
%

%
%
%
%
%
%
\section{Alternative formulation of $A$-pseudohermitian symmetries as ordinary (commuting) symmetries}
\label{app5}
Symmetry relations like \eqref{gs2} (for $A$ either linear or antilinear) may also be expressed as ordinary (commuting) symmetries, generalizing for scattering Hamiltonians the work in \cite{Mosta2002a,Mosta2002b,Mosta2002c}. In other words, for a Hamiltonian $H$ and a linear hermitian (antilinear hermitian) operator $A$ satisfying \eqref{gs2} we can find an antilinear (linear) operator $B$ that conmutes with $H$. In this appendix we explicitly construct the operators $B$ from the the Hamiltonian both for $A$ linear and antilinear in the first and second sections respectively.

Let us assume for now that besides $A$ (linear or antilinear) there exists an invertible and hermitian antilinear operator $\tau$ that also satisfies \eqref{gs2}. With $A$ and $\tau$ let us define the operator $B = A^{-1} \tau$ that will be antilinear (linear) for $A$ linear (antilinear). As defined, $B$ commutes with the Hamiltonian, because
\begin{eqnarray}
    B H &=& A^{-1}\tau H \nonumber \\
           &=& A^{-1} H\da \tau \nonumber \\
           &=& H A^{-1} \tau\nonumber \\
           &=& H B,
    \label{eq:HiddenSymmetry}
\end{eqnarray}
%
Note that $B$ is not generally Hermitian unless $\tau$ conmutes with $A^{-1}$.

The main task  to define  $B$ is to find the antilinear operator $\tau$ that satisfies \eqref{gs2}.
This can be achieved if the eigenvectors of the Hamiltonian and its adjoint form bases of the Hilbert space that are biorthonormal.
In \cite{Mosta2002c} the expression of $\tau$ for a discrete spectrum (with no degeneracy) is found as
\begin{equation}
    \tau_d \ket{\zeta} = \sum_n \braket{\zeta}{\phi_n} \ket{\phi_n},
    \label{eq:GenericTau}
\end{equation}
where the $d$ subscript indicates that the Hamiltonian has a discrete spectrum. The action of the operator in Eq. \eqref{eq:GenericTau} on a vector in an eigenspace amounts to complex conjugation of its coordinate
representation. $\tau_d$ is clearly antilinear, Hermitian (for antilinear operators hermicity is defined as $\braket{\chi}{\tau_d \zeta} = \braket{\zeta}{\tau_d \chi}$), and invertible. It can be checked that the relation $\tau_d H = H\da \tau_d$ is satisfied.

To generalize this to Hamiltonians whose spectrum includes a continuous part, we have to build an antilinear operator $\tau$ that acts in both the subspaces, ${\cal H}_d$ and ${\cal H}_c$, that are respectively spanned by the eigenfunctions associated with the discrete (point) and continuous spectra of the Hamiltonian. We propose to take $\tau= \tau_d + \tau_c$, where $\tau_{c/d}$ maps ${\cal{H}}_{c/d}$ to ${\cal H}_{c/d}$ and annihilates
${\cal{H}}_{d/c}$. Specifically, we take $\tau_d$ to be given by Eq. (\ref{eq:GenericTau}) with  $n$ denoting the eigenvectors of the Hamiltonian associated with the discrete part of the spectrum, and set
\begin{equation}
    \tau_c \ket{\zeta} = \int_{-\infty}^{\infty} dp \; \left[ C_+(p) \braket{\zeta}{\widehat{p}^+}\ket{\widehat{p}^+} + C_-(p)\braket{\zeta}{\widehat{p}^+}\ket{-\widehat{p}^+} \right],
    \label{eq:ContinuousTau}
\end{equation}
%
%
where $C_+(p)$ and $C_-(p)$ are complex coefficients. The operator in \eqref{eq:ContinuousTau} is clearly antilinear because of the antilinearity of the inner product with respect to its first argument. Hermicity of $\tau$ requires $C_-(p) = C_-(-p)$. When $\tau$ (which should satisfy Eq. \eqref{gs2}) acts on a a right scattering
eigenfunction, it must give a left eigenfunction with same energy. This is so because
\begin{eqnarray}
    H\da \tau \ket{p^+} &=& \tau H \ket{p^+}\nonumber\\
    &=& \tau E_p\ket{p^+}\nonumber\\
    &=& E_p \tau \ket{p^+}\nonumber\\
    &\big\Downarrow&\nonumber\\
    \tau\ket{p^+}&\propto&\ket{\widehat{p}^+}, \ket{-\widehat{p}^+},
\end{eqnarray}
%
where $E_p = \frac{p^2}{2m}$.
The condition that $\tau$ be invertible restricts the coefficients in Eq. (C3) further.
Consider the \textit{on shell} representation of $\tau_c$, $\braket{p^+}{\tau q^+} = \frac{|p|}{m}\delta(E_p-E_q) \mathsf{C}_{p,q}$, with $\mathsf{C}_{p,q} \equiv \delta_{p,q} C_+(q) + \delta_{p,-q} C_-(q)$, or in matrix form
\begin{equation}
\mathsf{C}(p)=
    \begin{pmatrix}
    C_+(p) & C_-(p) \\
    C_-(p) & C_+(-p)
    \end{pmatrix}.
\end{equation}
Since $\tau$ has to be invertible, $\mathsf{C}(p)$ must be invertible as well. This implies  $C_+(p)C_+(-p) - C_-(p)C_-(p) \neq 0$.

In the following sections we construct expressions for $B$, in section 1 for $A$ linear, and in section 2 for $A$ antilinear.

\subsection{$A$ linear\label{ContPseudoHermicity}}
In \cite{Mosta2002a,Mosta2002b,Mosta2002c} it is shown that pseudohermitian Hamiltonians, i.e. those satisfying \eqref{gs2} for $A=\eta$ with $\eta$ a hermitian and invertible linear operator, posses an energy spectrum whose complex eigenvalues come in complex-conjugate pairs. Moreover the eigenspaces associated with  the eigenvalues $E$ and $E^*$ have the same degeneracy and $\eta$ maps one to the other. Conversely, if the complex part of the spectrum of $H$ contains only complex-conjugate pairs, it can be shown that there exists an $\eta$ for which the Hamiltonian satisfies \eqref{gs2}.
These results hold for a general class of diagonalizable Hamiltonians with a discrete spectrum. For these Hamiltonians we can identify $\eta$ with
%
\begin{equation}
    \begin{split}
    \eta_{d} &= \sum_{n_0} \ketbra{\phi_{n_0}}{\phi_{n_0}}\\ &+ \sum_n \Big[\ketbra{\phi_{n_-}}{\phi_{n_+}} +  \ketbra{\phi_{n_+}}{\phi_{n_-}}\Big].
    \end{split}
    \label{eq:DiscreteEta}
\end{equation}
%
%
%
where the states $\ket{\psi_{n_0}}$ ($\ket{\phi_{n_0}}$) correspond to the right (left) eigenstates of $H$ with real energy $E_{n_0}$.
$\ket{\psi_{n+/n-}}$  (respectively $\ket{\phi_{n+/n-}}$) correspond to the right (respectively left) eigenvectors whose eigenvalue $E_{n+/n-}$  has a positive/negative imaginary part.
This gives $\eta_d \ket{\psi_{n_0}} = \ket{\phi_{n_0}}$, $\eta_d\ket{\psi_{n_+}} = \ket{\phi_{n_-}}$ and $\eta_d\ket{\psi_{n_-}} = \ket{\phi_{n_+}}$. Clearly $\eta_d$ is compatible with pseudohermicity since it maps right eigenvectors associated with the eigenvalue $E$ into left eigenvectors with eigenvalue $E^*$ and the pseudohermicity relation \eqref{gs2} is satisfied. To generalize \eqref{eq:DiscreteEta} for a scattering Hamiltonian we must add an additional term $\eta_c$ which acts on the subspace of scattering states and is compatible with the Hermiticity and invertibility of $\eta = \eta_d + \eta_c$. Since $\eta$ should transform the right scattering states into left ones in the same energy shell, $\eta_c \ket{p^+}$ should be a linear combination of both $\ket{\widehat{p}^+}$ and $\ket{-\widehat{p}^+}$. Accordingly, we propose
\begin{equation}
    \eta_c = \int_{-\infty}^\infty dp\; \left[\Lambda_+(p)\ketbra{\widehat{p}^+}{\widehat{p}^+} + \Lambda_-(p)\ketbra{-\widehat{p}^+}{\widehat{p}^+}\right],
    \label{eq:EtaContinuousPart}
\end{equation}
where $\Lambda_+(p), \Lambda_-(p)$ are complex coefficients depending on the momentum $p$. Hermicity of $\eta$ requires $\Lambda_+(p) \in \mathbb{R}$ and $\Lambda_-^*(p) =\Lambda_-(-p)$.

Since $\eta_c$ connects scattering states with the same energy it admits the following \textit{on-shell} representation, $\bra{q^+}\eta_c\ket{p^+} = \frac{|p|}{m}\delta(E_q-E_p) \mathsf{\Lambda}_{q,p}(p)$, with $\mathsf{\Lambda}_{q,p}(p) \equiv \delta_{q,p} \Lambda_+(p) + \delta_{q,-p} \Lambda_-(p)$, or in matrix form
\begin{equation}
    \mathsf{\Lambda}(p)
    =
    \left(
    \begin{matrix}
        \Lambda_+(p) & \Lambda_-^*(p) \\
        \Lambda_-(p) & \Lambda_+(-p)
    \end{matrix}
    \right).
    \label{eq:onShellEtaContinuous}
\end{equation}
%
Since $\eta$ has to be invertible, this implies that the determinant of $\mathsf{A}(p)$ should not vanish i.e. $\Lambda_+(p) \Lambda_+(-p) - \Lambda_-(p) \Lambda_-^*(p) \neq 0$. The inverse of $\eta$ is then $\eta^{-1} = \eta_d^{-1} + \eta_c^{-1}$ with
\begin{equation}
  \begin{split}
  \eta_d^{-1} &= \sum_{n_0} \ketbra{\psi_{n_0}}{\psi_{n_0}}\\
  &+\sum_{n} \left[\ketbra{\psi_{n_+}}{\psi_{n_-}} + \ketbra{\psi_{n_-}}{\psi_{n_+}}\right]
\end{split}
\end{equation}
\begin{equation}
    \eta_c^{-1} = \int_{-\infty}^\infty dp\; \left[\Lambda_+^{(-1)}(p)\ketbra{p^+}{p^+} + \Lambda_-^{(-1)}(p)\ketbra{-p^+}{p^+}\right],
\end{equation}
where the complex coefficients $\Lambda_\pm^{(-1)}(p)$ are taken from the inverse of $\mathsf{\Lambda}(p)$
\begin{equation}
  \mathsf{\Lambda}^{-1}(p)
  =
  \left(
  \begin{matrix}
      \Lambda_+^{(-1)}(p) & \Lambda_-^{(-1)*}(p) \\
      \Lambda_-^{(-1)}(p) & \Lambda_+^{(-1)}(-p)
  \end{matrix}
  \right).
  \label{eq:onShellEtaContinuous2}
\end{equation}
Using the orthogonality between the subspace of discrete (bound) and scattering states we find the final expression for $B$
\begin{equation}
  \begin{split}
  B\ket{\zeta} &= \eta_d^{-1}\tau_d\ket{\zeta}+  \eta_c^{-1}\tau_c\ket{\zeta}\\
  &=\sum_{n_0}\braket{\zeta}{\phi_{n_0}}\ket{\psi_{n_0}}\\
  &+\sum_{n}\braket{\zeta}{\phi_{n_+}}\ket{\psi_{n_-}}\\
  &+\sum_{n}\braket{\zeta}{\phi_{n_-}}\ket{\psi_{n_+}}\\
  &+\int_{-\infty}^{\infty}dp\,\braket{\zeta}{\widehat{p}^+}
  \left[ \tilde{C}_+(p)\ket{p^+} + \tilde{C}_-(p)\ket{-p^+} \right],
  \end{split}
\end{equation}
with $\tilde{C}_\pm(p) = C_+(p)\Lambda_\pm^{(-1)}(p)+C_-(p)\Lambda_\mp^{(-1)}(-p)$. Note that the resulting operator $B$ is antilinear.

\subsection{Pseudohermicity with antilinear operators}

In this section we will consider the case where the operator $A$ appearing in Eq. (\ref{gs2}) is antilinear.  In ref. \cite{Mosta2002c} this is called antipseudohermicity, but we will not use this terminology in order to avoid confusion with antihermicity ($H = - H\da$). The effect of $A$ on a right eigenvector of $H$ is to transform it  into its corresponding biorthonormal partner, i.e. the left eigenvector corresponding to the same energy
\begin{eqnarray}
    H\da A \ket{\psi_n} &=& A H \ket{\psi_n}\nonumber\\
    &=& A E_n\ket{\psi_n}\nonumber\\
    &=& E_n^* A \ket{\psi_n}\nonumber\\
    &\big\Downarrow&\nonumber\\
    A\ket{\psi_n}&\propto&\ket{\phi_n}.
\end{eqnarray}
$A$ also admits a decomposition similar to Eq. \eqref{eq:GenericTau}.
The Hamiltonian satisfies Eq. (\ref{gs2})
with respect to $\tau$. One can check that $A^{-1}\tau$ is a linear symmetry of the Hamiltonian,
$HA^{-1}\tau - A^{-1}\tau H=0$. The expansion of $A$ on the discrete and scattering basis is
\begin{widetext}
\begin{equation}
    A \ket{\xi} = \sum_n g_n \braket{\xi}{\phi_n} \ket{\phi_n} + \int dp \;  \braket{\xi}{\widehat{p}^+}   \left[ G_+(p) \ket{\widehat{p}^+} + G_-(p)\ket{-\widehat{p}^+} \right],
\end{equation}
\end{widetext}
with $A\ket{p^+\!}\!=\!G_+(p)\ket{\widehat{p}^+\!}+G_-(p)\ket{\!-\!\widehat{p}^+\!}$ and  $g_n = \bra{\psi_n}A {\psi_n}\ra$.
As examples we have found the expressions of $B = A^{-1}\tau$ for $A_T = \Theta$(time reversal) and $A_{PT} = \Pi\Theta$ (PT). In both cases we have $A_T^{-1}=A_T$ and $A_{PT}^{-1}=A_{PT}$.
\subsubsection{PT Symmetry}
The action of $B_{PT} = A_{PT}\tau$ on an arbitrary state is
\begin{widetext}
\begin{eqnarray}
    B_{PT}\ket{\zeta} &=& A_{PT}\tau\ket{\zeta}
\nonumber \\
                                   &=& A_{PT} \left\{ \sum_n \braket{\zeta}{\phi_n} \ket{\phi_n} + \int dp\; \left[ C_+(p) \braket{\zeta}{\widehat{p}^+}\ket{\widehat{p}^+} + C_-(p)\braket{\zeta}{\widehat{p}^+}\ket{-\widehat{p}^+} \right] \right\}.
\end{eqnarray}
\end{widetext}
Using $A_{PT} H = H\da A_{PT}$ and table \ref{tab:MollerOperatorSyms} we have $A_{PT} \ket{\widehat{p}^\pm} = \ket{\widehat{p}^\mp}$. Note that The ``-" right scattering states can be expressed in terms of the ``+" right scattering states as
\begin{eqnarray}
    \ket{\widehat{p}^-} &=& \int dq\; \ket{q^+}\braket{\widehat{q}^+}{\widehat{p}^-}\nonumber\\
    &=& \int dq\; \ket{q^+}\bra{q}\widehat{\Omega}_+\da \widehat{\Omega}_- \ket{p}\nonumber\\
    &=& \int dq\; \ket{q^+}\bra{q} \widehat{S}\da \ket{p}\nonumber\\
    &=& \ket{p^+}\bra{p} \widehat{\mathsf{S}}\da \ket{p} + \ket{-p^+}\bra{-p} \widehat{\mathsf{S}}\da \ket{p}.
\end{eqnarray}
With all this the final form of $B_{PT}$ is
\begin{eqnarray}
    B_{PT} &=&   \sum_n (g_n^*)^{-1} \ketbra{\psi_n}{\phi_n} \nonumber \\
                &+& \int dp\; \left[ \tilde{C}_+^*(p)\ketbra{p^+}{\widehat{p}^+} + \tilde{C}_-^*(p)\ketbra{-p^+}{\widehat{p}^+} \right], \nonumber \\
\label{C13}
\end{eqnarray}

%
with $C_{\pm}^*(p) = C_{\pm}^*(p) \bra{\pm p} \widehat{\mathsf{S}}\da \ket{\pm p} + C_{\mp}^*(p) \bra{\pm p} \widehat{\mathsf{S}}\da \ket{\mp p}$.

\subsubsection{Time-Reversal Symmetry}
For time-reversal symmetry,
\begin{widetext}
\begin{equation}
    B_{T}\ket{\zeta} = A_{T}\tau\ket{\zeta} = A_{T} \left\{ \sum_n  \braket{\zeta}{\phi_n} \ket{\phi_n} + \int dp\; \left[ C_+(p) \braket{\zeta}{\widehat{p}^+}\ket{\widehat{p}^+} + C_-(p)\braket{\zeta}{\widehat{p}^+}\ket{-\widehat{p}^+} \right] \right\}.
\end{equation}
\end{widetext}
Since the time-reversal operator satisfies the relation \eqref{gs2} with the Hamiltonian, Table \ref{tab:MollerOperatorSyms} implies $A_T\ket{\widehat{p}^\pm} = \ket{-\widehat{p}^\mp}$. The linear symmetry operator can be expressed as in Eq. (\ref{C13}) but in this case $\tilde{C}_{\pm}^*(p) = C_{\pm}^*(p) \bra{\pm p} \widehat{\mathsf{S}}\da \ket{\mp p} + C_{\mp}^*(p) \bra{\pm p} \widehat{\mathsf{S}}\da \ket{\pm p}$.

\bibliographystyle{apsrev4-1}
\bibliography{biblio.bib}

%
%

\end{document}